\documentclass[letter, twocolumn]{jpsj3_1}
\usepackage{txfonts}
\usepackage{textcomp}
\usepackage{booktabs}

\title{Maki Parameter and Upper Critical Field of the Heavy-Fermion Superconductor UBe$_{13}$}

\author{Yusei SHIMIZU$^{1}$\thanks{E-mail address: yusei@phys.sci.hokudai.ac.jp}, Yoichi IKEDA$^{1, 2}$, Takumi WAKABAYASHI$^{1}$, Yoshinori HAGA$^{3}$, Kenichi TENYA$^{4}$, \\
 Hiroyuki HIDAKA$^{1}$, Tatsuya YANAGISAWA$^{5}$, and Hiroshi AMITSUKA$^{1}$
}
\inst{$^{1}$Graduate School of Science, Hokkaido University, Sapporo 060-0810, Japan. \\
$^{2}$Graduate School of Natural Science and Technology, Okayama University, Okayama 700-8530, Japan. \\
$^{4}$Advanced Science Research Center, Japan Atomic Energy Agency, Tokai 319-1195, Japan. \\
$^{3}$Faculty of Education, Shinshu University, Nagano 390-8621, Japan. \\
$^{5}$Creative Research Institution, Hokkaido University, Sapporo 001-0021, Japan. 
}
\abst{We have performed low-temperature specific-heat measurements in magnetic fields for a single crystal UBe$_{13}$. 
It has been observed that our sample exhibits a superconducting transition  at an intermediate temperature ($T_{\mathrm{c}}$ $\sim$ 0.81 K)
 between previously reported values for two variant samples called H type and L type.
The specific heat $C(T)$ of our sample shows a $T^{3}$ behavior
 below $\sim$ 0.7 $T_{\mathrm{c}}$, 
which is similar to the behavior of the H-type sample, suggesting the existence of point nodes in the superconducting gap function. 
We have obtained the upper-critical-field curves $H_{\mathrm{c2}}$ for the [001], [110], and [111] crystal axes, 
which show no anisotropy at least down to the lowest measured temperature of 0.5 K.
We  have also derived  the Maki parameter $\kappa_{2}$, 
and 
it has been revealed that the $\kappa_{2}$ steeply decreases isotropically upon cooling just below $T_{\mathrm{c}}$. 
Paramagnetic effects and the symmetry of Cooper pairing of UBe$_{13}$ are discussed. 
}
\kword{Unconventional Superconductor, Heavy Fermion, Paramagnetic Effect}

\begin{document}
\maketitle


The heavy-fermion compound UBe$_{13}$ has attracted much attention  
 since the discovery of  unconventional superconductivity 27 years ago \cite{Ott1983}. 
UBe$_{13}$ has a cubic structure with the space group $O_{h}$$^{6}$(Fm$-$3c), and exhibits a superconducting (SC) transition 
at around 0.85 K with a large specific-heat jump $\Delta C$.
%
Despite prolonged scientific efforts to unravel the superconductivity of this material, 
many open questions remain, in particular concerning the symmetry of Cooper pairing and the unusual SC phase diagram
 \cite{Rauchschwalbe1985, Maple1985}. 
%

As for the SC gap structure, 
which is related to the parity of Cooper pairing, the temperature dependence of 
  low-$T$ specific heat $C(T)$ \cite{Ott1984} and  magnetic field penetration depth $\lambda(T)$
 \cite{Einzel1986} suggest the presence of  point nodes, whereas nuclear magnetic resonance (NMR) spin-relaxation rate $1/T_{1}$ 
 \cite{Maclaughlin}
and ultrasonic attenuation $\alpha(T)$
\cite{BGolding}
 suggest  a line node.
A $p$-wave Anderson-Brinkman-Morel (ABM) state, which is identified as the A phase in superfluid $^{3}$He 
\cite{Leggett1975}, 
has been proposed for the SC of UBe$_{13}$ from the $T^{3}$-like behavior observed in the low-$T$ specific heat
 \cite{Ott1984}.
An early NMR study also suggests the occurrence of the  ABM state,  because the Knight shift does not decrease below $T_{\mathrm{c}}$
 \cite{Tien1989}. 
 However, recent muon spin-rotation ($\mu$SR) measurements performed in field,  $H$ $||$ [001],  have
  revealed that the muon Knight shift shows a clear decrease below $T_{\mathrm{c}}$
 \cite{Sonier2003}.
Furthermore, 
 a recent NMR study by Tou $et$ $al$.
 has suggested that the Knight shift decreases below $T^{*}$ ( $\sim$ 0.7 K $<$ $T_{\mathrm{c}}$) for [001], whereas 
 increases  below $T^{*}$ for [111]
 \cite{Tou2010}. 
Therefore, no consensus has been achieved regarding the behavior of Knight shift below $T_{\mathrm{c}}$ as well as the SC-gap structure.
 Namely, it is still debated whether the Cooper-pairing state of UBe$_{13}$ is triplet or singlet.

In order to obtain further information about the parity of Cooper pairing, 
 it is important to investigate the presence (or absence) of paramagnetic effects.
The value of $H_{\mathrm{c2}}$ estimated experimentally by extrapolation to $T = 0$ for UBe$_{13}$ is about 80-130 kOe, which is much larger than the Pauli limit $H_{\mathrm{P}}$($T$=0) of 18.4 $T_{\mathrm{c}}$ kOe
 ($\sim$ 16.6 kOe for $T_{\mathrm{c}}=0.9$ K).
The paramagnetic effects might be thus considered to be absent in UBe$_{13}$.
However, this expression of $H_{\mathrm{P}}$ is reliable only when 
a spin-orbit scattering by magnetic impurities can be neglected.
Therefore,  this large $H_{\mathrm{c2}}$ compared with the $H_{\mathrm{P}}$ is insufficient to conclude that there is no paramagnetic effect 
  in UBe$_{13}$.

Unconventional superconductors often exhibit anisotropy of  $H_{\mathrm{c2}}(T)$, which could reflect the symmetry of 
SC order parameter \cite{Gorkov1984} 
as well as anisotropy of effective mass of electron. 
%
%
Aliev $et$ $al$. have reported that a single crystal of  UBe$_{13}$ shows an anisotropy of $H_{\mathrm{c2}}$ between [001] and [110] directions
 \cite{Aliev1991},  
 while Signore $et$ $al$. have observed no anisotropy 
 \cite{Signore1995}.
This discrepancy is probably due to a sample-quality difference; 
 to our knowledge, however,  there has been no experimental study on the anisotropy of $H_{\mathrm{c2}}$ for UBe$_{13}$, except these two reports.
The further study using another single crystal will thus be useful and necessary.

In the present work, 
we have performed  low-$T$ specific heat measurements on a single crystal of UBe$_{13}$ under magnetic fields applied for  three different crystal axes, 
 and investigated the anisotropy of  $H_{\mathrm{c2}}$.
We have also studied  the  temperature dependence and the anisotropy of the Maki parameter $\kappa_{2}$,  which 
reflects the paramagnetic effects of Cooper pairing.





A single crystal  UBe$_{13}$ was grown by an Al-flux method described in ref. 15.
Its dimensions were $\sim$ 1.8 $\times$ 1.9 $\times$ 0.8 mm$^3$
 and weight was 6.6 mg.
The sample orientation was checked  by X-ray Laue methods. 
The specific-heat measurements  were performed by a standard 
semi-adiabatic heat-pulse method in zero magnetic field and by  a thermal relaxation method in magnetic fields,  
using  $^{3}$He refrigerators. 
The magnetic field was applied up to 4.5 T,  along the three different cubic crystal axes: [001], [110], and [111].



Figure 1 shows the temperature dependence of the zero-field specific heat of UBe$_{13}$ above  0.3 K.
The peak-top temperature $T^{\mathrm{top}}_{\mathrm{c}}$ of the specific-heat jump  is estimated to be $\sim$ 0.79 K.  
From the entropy conservation for the SC transition,  on the other hand, 
$T_{\mathrm{c}}$ is estimated to be $\sim$ 0.81 K.
Recently, it has been reported that there are two variants in pure UBe$_{13}$, 
 called H (high) and L (low) type  
\cite{Langhammer1998}. 
An H-type UBe$_{13}$ exhibits a SC transition at $\sim$ 0.85-0.9 K, 
 and has a broad anomaly at $T_{\mathrm{L}} \sim 0.7$ K in thermal expansion under zero magnetic field. 
This anomaly cannot be seen  in the temperature dependence of the specific heat at zero field,
 but in field,
  the isothermal specific heat curves $C(B)$ show a broad anomaly at a 
  corresponding field called  $B^{*}(T_{\mathrm{L}})$
 \cite{Kromer1998}.  
On the other hand, an L-type UBe$_{13}$ has a lower  $T_{\mathrm{c}}$ of $\sim$ 0.75 K, 
 and it exhibits an obvious anomaly named $T_{\mathrm{A}}$ below $T_{\mathrm{c}}$
  in the temperature dependence of the zero-field specific heat.   
It has been found that $T_{\mathrm{A}}$ corresponds to the thermal-expansion anomaly $T_{\mathrm{L}}$
 \cite{Langhammer1998}. 

In the present study, we could not detect any anomalies in the specific heat 
below $T_{\mathrm{c}}$  within  the experimental accuracy.
It should also be noted that the specific-heat jump $\Delta C$ at $T_{\mathrm{c}}$ of our sample is larger than those of L-type and H-type samples.
It is considered that UBe$_{13}$ is a strong-coupling superconductor 
because a dimensionless value,  $\Delta C/ C_{\mathrm{n}}|_{T=T_{\mathrm{c}}}(\equiv \delta C)$ are considerably larger than 1.43,  which is expected  from the weak-coupling BCS theory.
The  $\delta C$ value of an H-type sample is reported to be $\sim$ 2.6 and a L-type sample  $\sim$ 2.0
\cite{Langhammer1998, Kromer1998}.  
on the other hand, our sample shows $\delta C$ of $2.9 \pm 0.3$, if estimated by the entropy conservation analysis, 
 and this value is close to the value of the H-type sample.
Hence, 
 the  properties seen in the specific heat of our sample is similar to those of the H-type sample 
 rather than the L-type sample,  although the $T_{\mathrm{c}}$  is different from the values of each type. 
%
%
%
\begin{figure}
\includegraphics[width=8cm]{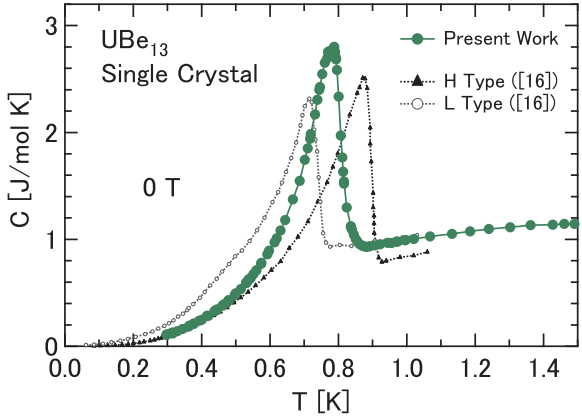}
\caption{\label{label} The temperature dependence of the specific heat of the single crystalline UBe$_{13}$ in zero field. 
Considering the entropy conservation at second-order transition,  
$T_{\mathrm{c}}$ is estimated to be 0.81 K. 
Data of H-type and L-type samples from ref. 16 are also plotted for comparison.
}
\end{figure}

\begin{figure}
\includegraphics[width=7.6cm]{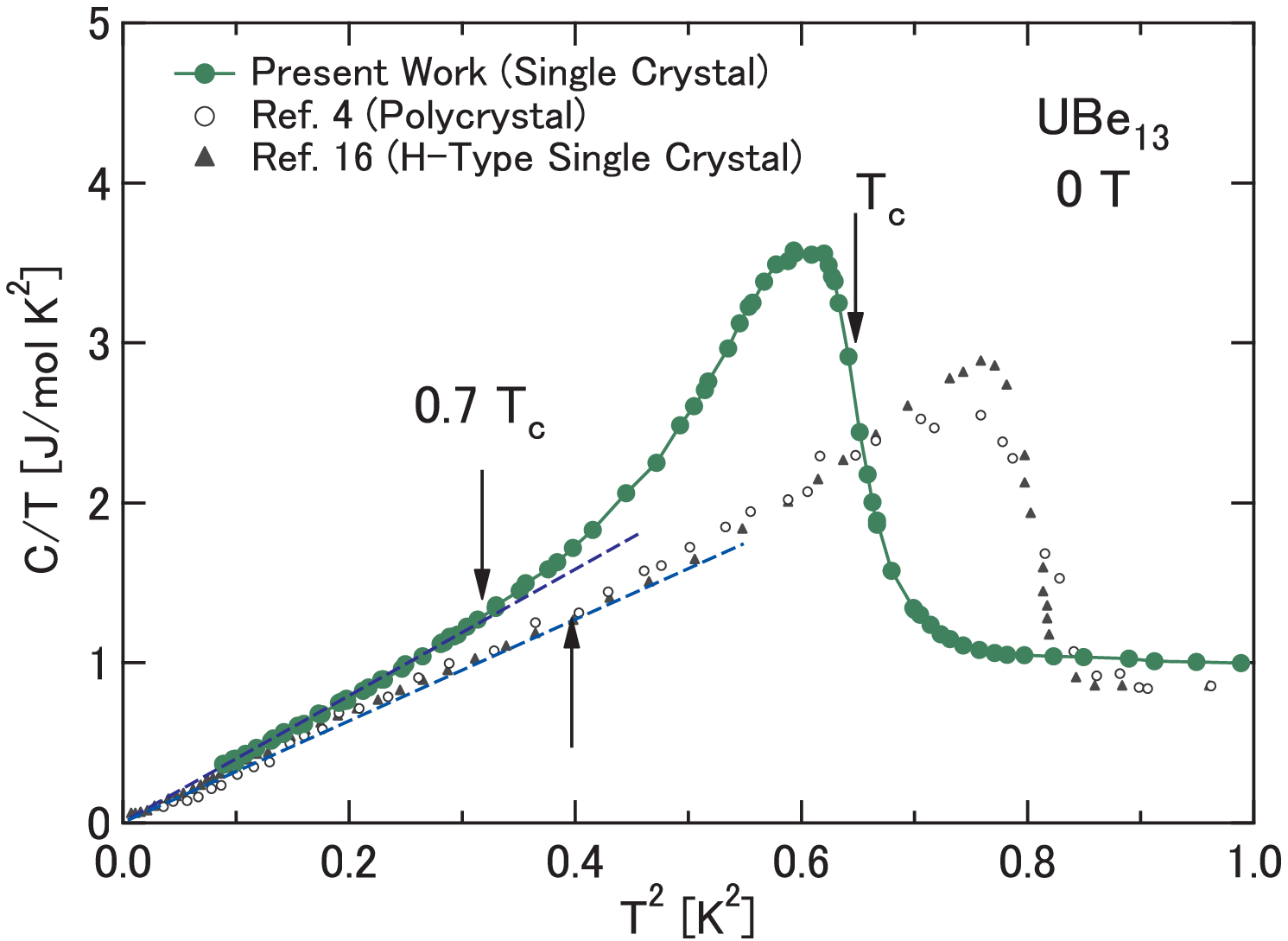}
\caption{\label{label} $C(T)/T$ vs $T^2$ plot of UBe$_{13}$ in zero field. 
Data from refs. 4 and 16 are also plotted for comparison. 
These  $C(T)$ data roughly obey the relation $C(T)$ $\propto$ $T^3$ in the temperature range below $\sim$ 0.7 $T_{\mathrm{c}}$, 
 which is indicated by arrows. Broken lines are guides to the eye.
}
\includegraphics[width=7.7cm]{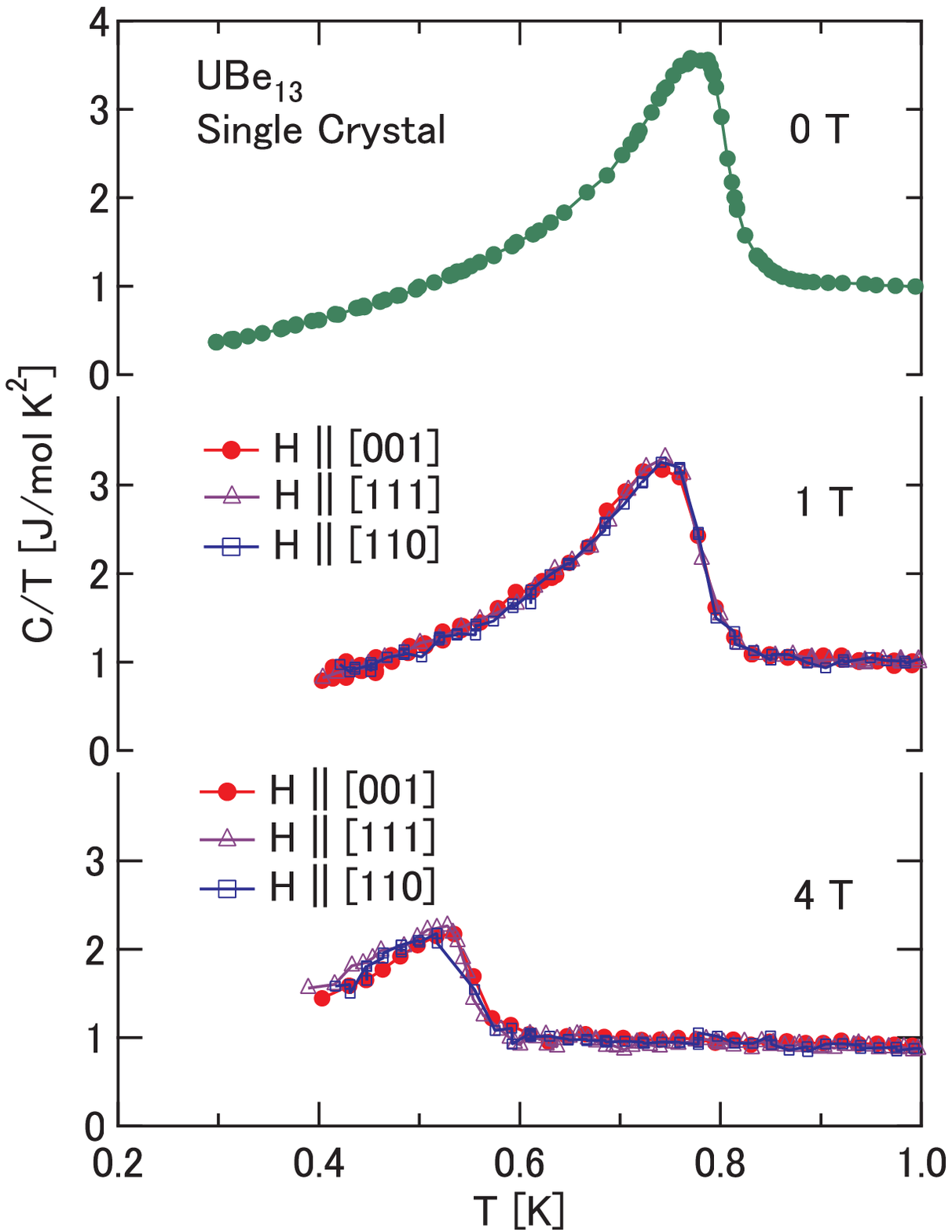}
\caption{\label{label} $C(T)/T$ of the single crystalline UBe$_{13}$ in 0, 1, and 4 T for three different crystal axes. }
\end{figure}


Figure 2 shows a  $C(T)/T$ versus $T^2$ plot of UBe$_{13}$ of our experimental results in zero field. 
The data reported for  an H-type polycrystal in  ref. 4 and 
 an H-type single crystal in ref. 16 are also plotted.
Although $T_{\mathrm{c}}$ is different, our sample and the H-type sample both reveal the $T^{3}$ power-law temperature dependence of $C(T)$ 
 in the low temperature range $T$ $<$ 0.7 $T_{\mathrm{c}}$.
This behavior supports the proposal by Ott $et$ $al$. that there are point nodes in the SC  gap  of UBe$_{13}$.
As mentioned above, however, the correspondence  between the $C(T)$ behavior and the 
microscopic observation associated with the SC gap is not clear in this system, because of its delicate dependence on sample quality. 
Therefore, in order to establish a reliable specification of the SC gap symmetry of UBe$_{13}$, it should be necessary to investigate the low-$T$ 
quasiparticle excitations by measuring various quantities on a same-quality single crystal. 
%

Figure 3 shows the temperature dependence of the  specific heat $C(T)/T$ measured at the fields 0, 1,  and 4 T for the directions: [001], [110], and [111].  
There is no significant anisotropy in the magnitude of $C(T)$ and $T_{\mathrm{c}}$ within the experimental error.
%
%
Figure 4 shows the results of  $H_{\mathrm{c2}}$ for [001] and $H_{\mathrm{c2}}^{\mathrm{top}}$ for [001], [110], and [111], 
 where $H_{\mathrm{c2}}$ and $H_{\mathrm{c2}}^{\mathrm{top}}$ are the upper critical fields estimated from $T_{\mathrm{c}}(H)$ and $T_{\mathrm{c}}^{\mathrm{top}}(H)$, 
 respectively.
The upper critical field obtained by our previous DC magnetization measurements for the same sample   
 \cite{YShimizu2011} is also plotted as  $H_{\mathrm{c2}}^{\mathrm{mag}}$.
%
%
\begin{figure}[!htb]
\includegraphics[width=8cm]{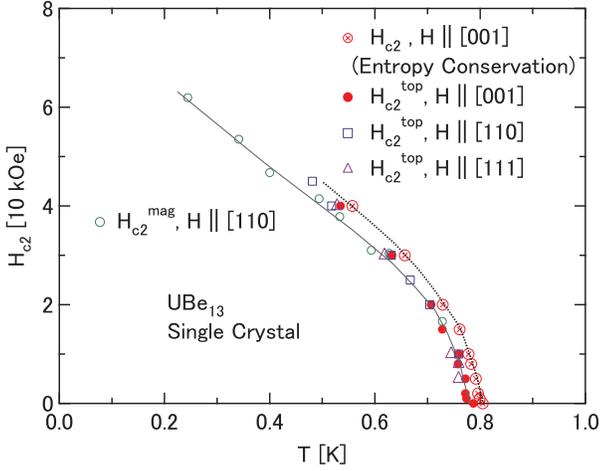}
\caption{\label{label} 
The upper critical field $H_{\mathrm{c2}}$ and $H_{\mathrm{c2}}^{\mathrm{top}}$ of UBe$_{13}$ obtained from specific-heat measurements.
 As for the  $H_{\mathrm{c2}}^{\mathrm{top}}$, the results for three different crystal axes ([001], [110], and  [111]) are plotted.
 The solid and the dotted lines are guides to the eye.   
The results of previous DC magnetization measurements from ref. 18 are also plotted.  
 }
\end{figure}
%
%
$H_{\mathrm{c2}}^{\mathrm{top}}$ is in good agreement with $H_{\mathrm{c2}}^{\mathrm{mag}}$, 
while  $H_{\mathrm{c2}}$, which is obtained by considering entropy conservation at $T_{\mathrm{c}}$,  
 is slightly larger  than $H_{\mathrm{c2}}^{\mathrm{top}}$ and  $H_{\mathrm{c2}}^{\mathrm{mag}}$.
%
%
Previously, 
Aliev $et$ $al$. have reported that $H_{\mathrm{c2}}$ for the field direction [001] is larger than that for [110], 
and their difference increases to $\sim$ 300 Oe at $\sim$ 0.7 $T_{\mathrm{c}}$ 
 \cite{Aliev1991}.
Since $T_{\mathrm{c}}$ of our sample is 0.81 K, 
 the similar magnitude of  anisotropy is  expected to be observed at $\sim$ 0.5 K. 
%
%
%
However, as seen in Fig. 3 and 4, the specific heat of our sample does not show any evidence for the anisotropy of $H_{\mathrm{c2}}$, at least, down to 0.5 K. 
It will be very important to investigate whether the anisotropy appears in the $H_{\mathrm{c2}}$ curves at lower temperatures.  


In general, 
 the Maki parameter $\kappa_{2}$
can be obtained from 
the slope of  a magnetization curve just below $H_{\mathrm{c2}}$ by using the relation (in cgs units):  
$
(dM^{\mathrm{SC}}_{\mathrm{eq}}/dH)_{H_{\mathrm{c2}}} 
= 1/4 \pi (2 \kappa^{2}_{2} - 1 ) \beta_{\mathrm{A}}, 
$
%
where the subscript `eq' denotes a thermal equilibrium process,  $M^{\mathrm{SC}}_{\mathrm{eq}}(H)$  a diamagnetic contribution of superconductivity,  and 
$\beta_{\mathrm{A}} (= 1.16$)  a geometric constant for triangular vortex lattice
 \cite{Serin, Saint-James}.
On the other hand,  Ehrenfest's relation for a second-order phase transition relates   
the slope of magnetization curve near $H_{\mathrm{c2}}$ to the specific-heat jump $\Delta C$
 as follows
 \cite{Goodman1962, Saint-James}:   
\begin{equation}
 \left( \frac{\Delta C}{T}\right)_{T_{\mathrm{c}}(H)} = \left( \frac{dH_{\mathrm{c2}}}{dT} \right)^2_{T_{\mathrm{c}}(H)} \left( \frac{dM^{\mathrm{ SC}}_{\mathrm{eq}}}{dH} \right)_{H_{\mathrm{c2}}}.
\end{equation}   
%
Consequently, 
 one can obtain $\kappa_{2}$ also from the observed values of $dH_{\mathrm{c2}} /dT$ and 
 $\Delta C$ at $T_{\mathrm{c}}(H)$ in magnetic fields as follows:
\begin{equation}
 \left( \frac{\Delta C}{T} \right)_{T_{\mathrm{c}}(H)}  = \left( \frac{d H_{\mathrm{c2}}}{dT} \right)^{2} \frac{1}{4 \pi (2 \kappa^{2}_{2} -1 ) \beta_{A}}.
\end{equation} 
In order to perform this latter procedure, 
we converted the specific heat per mole [J/mol K] to the value per unit volume [erg/cm$^{3}$K] by using 
  the  density and molecular mass of  UBe$_{13}$: $d$ = 4.36 g/cm$^{3}$ and $M = 355.187$, respectively.  
%
%
%
%

\begin{figure}
\includegraphics[width=7.9cm]{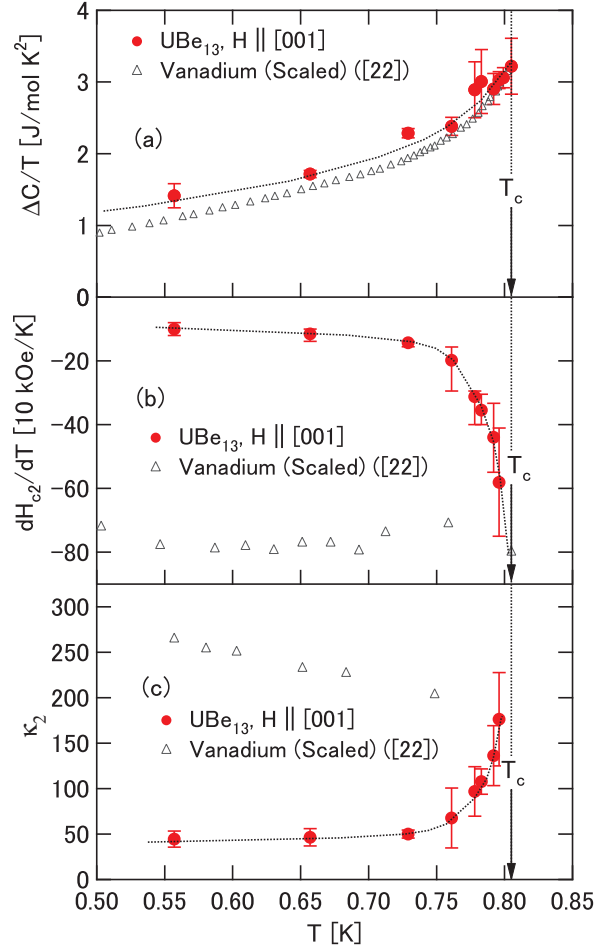}
\caption{\label{label} The temperature dependence of (a) $\Delta C$,  (b) $dH_{\mathrm{c2}}/dT$, and (c)  $\kappa_{2}$ for $H$ $||$ [001] of UBe$_{13}$. 
The dotted lines are guides to the eye.
The experimental results of vanadium  \cite{Radebaugh1966}, scaled to the magnitudes of UBe$_{13}$ at zero field, are also plotted for comparison.  
}
\end{figure}
%
%
%

Figure 5 (a), (b), and (c) show  the obtained temperature dependence of $\Delta C/T$, $dH_{\mathrm{c2}}/dT$,   
and $\kappa_{2}$, respectively, for $H$ $||$ [001].
$\kappa_{2}$ of our sample decreases steeply with decreasing temperature below $T_{\mathrm{c}}$, and becomes constant below $\sim$ 0.88 $T_{\mathrm{c}}$.
According to Eq. (2), $\kappa_{2}$ is almost inversely proportional to $(\Delta C/T)^{1/2}$, and directly proportional to 
$|dH_{\mathrm{c2}}/dT|$ in the case of $\kappa_{2} >> 1$.
The decrease in $\Delta C/T$ thus contributes to increasing $\kappa_{2}$, 
while that in  $|dH_{\mathrm{c2}}/dT|$ reduces $\kappa_{2}$.
In the present case, as temperature is lowered from $T_{\mathrm{c}}$ to $\sim$ 0.5 K, 
$|dH_{\mathrm{c2}}/dT|$ decreases to $\sim$ 16 $\%$, 
 whereas $(\Delta C/T)^{1/2}$ decreases to $\sim$ 66 $\%$.
The observed reduction of $\kappa_{2}$ with the non-linear $T$ variation in UBe$_{13}$ can 
thus mainly be attributed to the behavior of $dH_{\mathrm{c2}}/dT$.
Since there is no anisotropy in $\Delta C$ and $H_{\mathrm{c2}}$, 
  $\kappa_{2}(T)$ shows no anisotropy as well.

For the sake of comparison, we also plot in Fig. 5 the results of a type-II superconductor, vanadium (V), quoted from ref. 22.
The magnitude of $T_{\mathrm{c}}$, $\Delta C/T|_{T = T_{\mathrm{c}}}$, $dH_{\mathrm{c2}}/dT|_{T = T_{\mathrm{c}}}$,
 and $\kappa_{2}(T_{\mathrm{c}})$ $(= \kappa_{\mathrm{GL}} \equiv \lambda/\xi)$ of V are scaled to those of UBe$_{13}$ at zero field, 
 where $\lambda$ and $\xi$ denote magnetic penetration depth and correlation length, respectively.
The relative changes of $\Delta C/T$ of V is similar to that of  UBe$_{13}$,
 while $dH_{\mathrm{c2}}/dT$ of V is nearly independent of temperature, 
 providing a remarkable contrast with the behavior of UBe$_{13}$.
As a result, $\kappa_{2}$ of V just shows a monotonous increase upon cooling. 
From this comparison, we strongly suggest 
the peculiarity of the field distraction process of the SC state in UBe$_{13}$; 
 i. e., a strong bending effect of the $H_{\mathrm{c2}}$ curve appears to be present at $\sim$ 0.9 $T_{\mathrm{c}}$. 
We should also note that the absolute value of $\kappa_{\mathrm{GL}}$ for UBe$_{13}$ is 
 roughly of order $10^{2}$, which is two orders of magnitude larger than that of V ($\sim$ 0.85).
%
%
%
%
%
%
%
%

The decrease of  $\kappa_{2}$  upon cooling could reflect a paramagnetic effect.  
Generally, energy due to spin susceptibility $\chi_{\mathrm{s}}H^{2}/2$ of a heavy-fermion system 
 is large and comparable with the  condensation energy of SC state:   $\chi_{\mathrm{s}}H^{2}/2 \sim H_{\mathrm{c}}^{2}/8\pi$.
In this case, when two  antiparallel-spin electrons form a Cooper pair, 
 the  electrons could change from the SC state to the spin-polarized normal state in a high field ($\sim H_{\mathrm{P}}$)
 to gain the magnetic energy of the system.  
Consequently, the $H_{\mathrm{c2}}(T)$ in high-field region
 is suppressed (paramagnetic effect)
 \cite{Werthamer1966}.
Since the slope of magnetization curve $dM^{\mathrm{SC}}_{\mathrm{eq}}(H)/dH$ near $H_{\mathrm{c2}}$ becomes larger at low temperatures,  $\kappa_{2}(T)$ 
decreases upon cooling. 
In the case of spin-singlet pairing, the paramagnetic effect
 should be present for all directions, and thus  $\kappa_{2}(T)$
 isotropically decreases by cooling in fields.
For example, the heavy-fermion superconductor CeCoIn$_{5}$, which is considered to have  an even-parity pairing,  
reveals a clear decrease of $\kappa_{2}$ regardless of the field direction  
 \cite{Ikeda2001}.
In the case of spin-triplet pairing, on the other hand, if spin-orbit coupling is negligibly small, 
 the paramagnetic effect should be absent for all directions,  
 since the total spin of Cooper pair can follow the field direction without conflicting with the SC condensation. 
In this case, the $\kappa_{2}(T)$ in general does not decrease but increases  
upon cooling.
However,  if the spin-orbit coupling is not negligible, it may lead to  
the paramagnetic effect:   
 if  
a projection of a total spin of a Cooper pair perpendicular to the magnetic field  is not zero, and locked on one direction of a crystal axis with strong spin-orbit coupling, 
the paramagnetic effect should be present. 
%

%
Our experimental results  suggest that 
 the paramagnetic effect isotropically occurs in UBe$_{13}$, and  
  this does not conflict with  the behavior expected for the  spin-singlet pairing. 
An odd-parity pairing , however, might also be possible, if it leads to some isotropic 
paramagnetic effect at least down to 0.5 K ($\sim$ 0.6 $T_{\mathrm{c}}$). 
Such an  isotropic paramagnetic effect may occur in a triplet 
state, 
 with strong spin-orbit coupling, 
 which belongs to a one-dimensional representation of the cubic group: 
the SC gap function of this state is constant over the Fermi surface, and is similar to Balian-Werthamer state of $^{3}$He superfluidity.
However, this one-dimensional triplet SC state cannot explain the existence of  point nodes in the SC gap of this material.
As for another possible explanation for the decrease in  $\kappa_{2}(T)$ of UBe$_{13}$, 
 we wish to point out the effect of  spin-orbit scattering by magnetic impurities.
Maki has numerically shown that  $\kappa_{2}$ of a type-II superconductor could decrease upon cooling 
 in the presence of  strong spin-orbit scattering in a dirty limit 
\cite{Maki1966}.
If   there are  5$f$ electrons that remain the normal state even in the SC phase, they  may act as the  magnetic impurities 
forming a scattering potential for the SC electrons.
Our present data  do not rule out the possibility  
 that   such spin-orbit scattering occurs frequently in UBe$_{13}$.   
%
%
%
In order to elucidate the origin of 
the decrease in the $\kappa_{2}(T)$ of this system,    
 it will  be  very crucial  to investigate  the temperature dependence and the anisotropy of the $\kappa_{2}$ at lower temperatures.
For this purpose,  DC magnetization measurements below 0.5 K are now in progress.

%
In conclusion, we have performed the low-$T$ specific heat measurements in magnetic fields on a single crystal of UBe$_{13}$.
The specific heat $C(T)$ of our sample obeys $T^3$ dependence in the temperature range of $T$ $<$ 0.7 $T_{\mathrm{c}}$, 
  which is similar to the behavior of the H-type samples \cite{Ott1984}, 
 and suggests  the presence of point nodes in the SC gap of this material.     
%
%
%
%
%
We have also derived the Maki parameter $\kappa_{2}(T)$ as well as  the upper critical field $H_{\mathrm{c2}}$
 down to 0.5 K for the three directions of [001], [110], and [111].
No anisotropy has been observed in the $\kappa_{2}(T)$ and $H_{\mathrm{c2}}(T)$ curves within the experimental accuracy.
We have shown that the $\kappa_{2}$ reveals a clear decrease upon cooling for all the directions.
The experimental results suggest the presence of paramagnetic effect, which is isotropic to the investigated crystal axes, 
 or of the effect of strong spin-orbit scattering due to magnetic impurities in UBe$_{13}$, both at least,  down to 0.5 K. 
In order to obtain  further understanding of  this unusual 
SC state, it is needed to investigate
 the behavior of  $\kappa_{2}$ and  $H_{\mathrm{c2}}$ at lower temperatures using the same sample.

We thank H. Tou, H. Harima, and H. Kusunose for valuable discussions.
The present work was supported by Grant-in-Aid for Scientific Research on Innovative Areas `Heavy Electrons' (20102002) and Scientific Research B (19340086),
 and S (20224015) from the Ministry of Education, Culture, Sports, Science and Technology, Japan.     


\end{document}